\documentclass[aps,prb,twocolumn,groupedaddress,floatfix,showpacs,showkeys]{revtex4}
\usepackage[dvips]{graphicx,color}

\newcommand{\micron}{$\mu$m}

\newcommand{\pgnfigure}[2]{\begin{figure}\includegraphics[height=6cm,clip=true]
{#1.eps}\caption{\label{#1}#2}\end{figure}}
\newcommand{\pgnfigurestwo}[3]{\begin{figure}\includegraphics[height=6cm,clip=true]
{#1.eps} \par \par\includegraphics[height=6cm,clip=true]{#2.eps}
\caption{\label{#1}#3}\end{figure}}
\newcommand{\pgnfiguresthree}[4]{\begin{figure}\includegraphics[height=5.5cm,clip=true]
{#1.eps} \par \par
\includegraphics[height=5.5cm,clip=true]{#2.eps} \par \par
\includegraphics[height=5.5cm,clip=true]{#3.eps}
\caption{\label{#1}#4}\end{figure}}

\begin{document}


\title{Unconventional resistivity at the border of metallic antiferromagnetism in NiS$_2$}



\author{P. G. Niklowitz}
\email[e-mail: ]{philipp.niklowitz@frm2.tum.de}
\altaffiliation[present address: ]{Physik Department E21,
Technische Universit\"at M\"unchen, 85747 Garching, Germany}
\author{M. J. Steiner}
\author{G. G. Lonzarich}
\affiliation{Cavendish Laboratory, University of Cambridge,
Madingley Road, Cambridge CB3 0HE, UK}
\author{D. Braithwaite}
\author{G. Knebel}
\author{J. Flouquet}
\affiliation{D\'{e}partement de Recherche Fondamentale sur la
Mati\`{e}re Condens\'{e}e, SPSMS,CEA Grenoble, 38054 Grenoble
Cedex 9, France}
\author{J. A. Wilson}
\affiliation{H. H. Wills Physics Laboratory, University of
Bristol, Bristol BS8 1TL, UK}


\date{\today}

\begin{abstract}

We report low-temperature and high-pressure measurements of the
electrical resistivity $\rho(T)$ of the antiferromagnetic compound
NiS$_2$ in its high-pressure metallic state. The form of $\rho(T)$
suggests that metallic antiferromagnetism in NiS$_2$ is quenched
at a critical pressure $p_c=76\pm 5$~kbar. Near $p_c$ the
temperature variation of $\rho(T)$ is similar to that observed in
NiS$_{2-x}$Se$_x$ near the critical composition $x=1$ where the
N\'eel temperature vanishes at ambient pressure. In both cases
$\rho(T)$ varies approximately as $T^{1.5}$ over a wide range
below 100~K. However, on closer analysis the resistivity exponent
in NiS$_2$ exhibits an undulating variation with temperature not
seen in NiSSe ($x=1$). This difference in behaviour may be due to
the effects of spin-fluctuation scattering of charge carriers on
cold and hot spots of the Fermi surface in the presence of
quenched disorder, which is higher in NiSSe than in stoichiometric
NiS$_2$.

\end{abstract}


\maketitle

\section{Introduction}

The electronic properties of metals on the border of magnetic
phase transitions at low temperatures are often found to exhibit
temperature dependences that differ from the predictions of Fermi
liquid theory.  Early attempts to explore such non-Fermi liquid
behaviour have been based on mean-field treatments of the effects
of enhanced magnetic fluctuations, as in the self-consistent
renormalization (SCR) model.\cite{her76a,mil93a,mor85a,lon97a}

In a recent work the prediction of this model for the temperature
dependence of the electrical resistivity $\rho(T)$ was tested in a
simple cubic d-metal, Ni$_3$Al, at high pressures near to the
critical pressure where ferromagnetism is
suppressed.\cite{nik04d}\ The $T^{5/3}$ temperature dependence of
the resistivity seen in Ni$_3$Al and other related systems, where
the magnetic correlation wavevector $\kappa(T)$ is small, appears
to be largely consistent with the SCR model.\cite{mat68a}\ In the
idealized limit $\kappa\rightarrow 0$ at $T\rightarrow 0$, the SCR
model predicts that in 3D the quasiparticle scattering rate
$\tau_{qp}^{-1}$ varies linearly with the quasiparticle excitation
energy, rather than quadratically as in the standard Fermi liquid
picture. This form of $\tau_{qp}^{-1}$ is similar to that of the
marginal Fermi liquid model,\cite{var89a,hol73a,bay91a}\ which is
normally associated with a linear temperature dependence of the
resistivity. However, at the border of ferromagnetism the relevant
fluctuations responsible for quasiparticle scattering are of long
wavelength and thus are ineffective in reducing the current. This
leads to a transport relaxation rate $\tau_{tr}^{-1}$ that differs
from $\tau_{qp}^{-1}$ and varies not as $T$, but as $T^{5/3}$

\pgnfigure{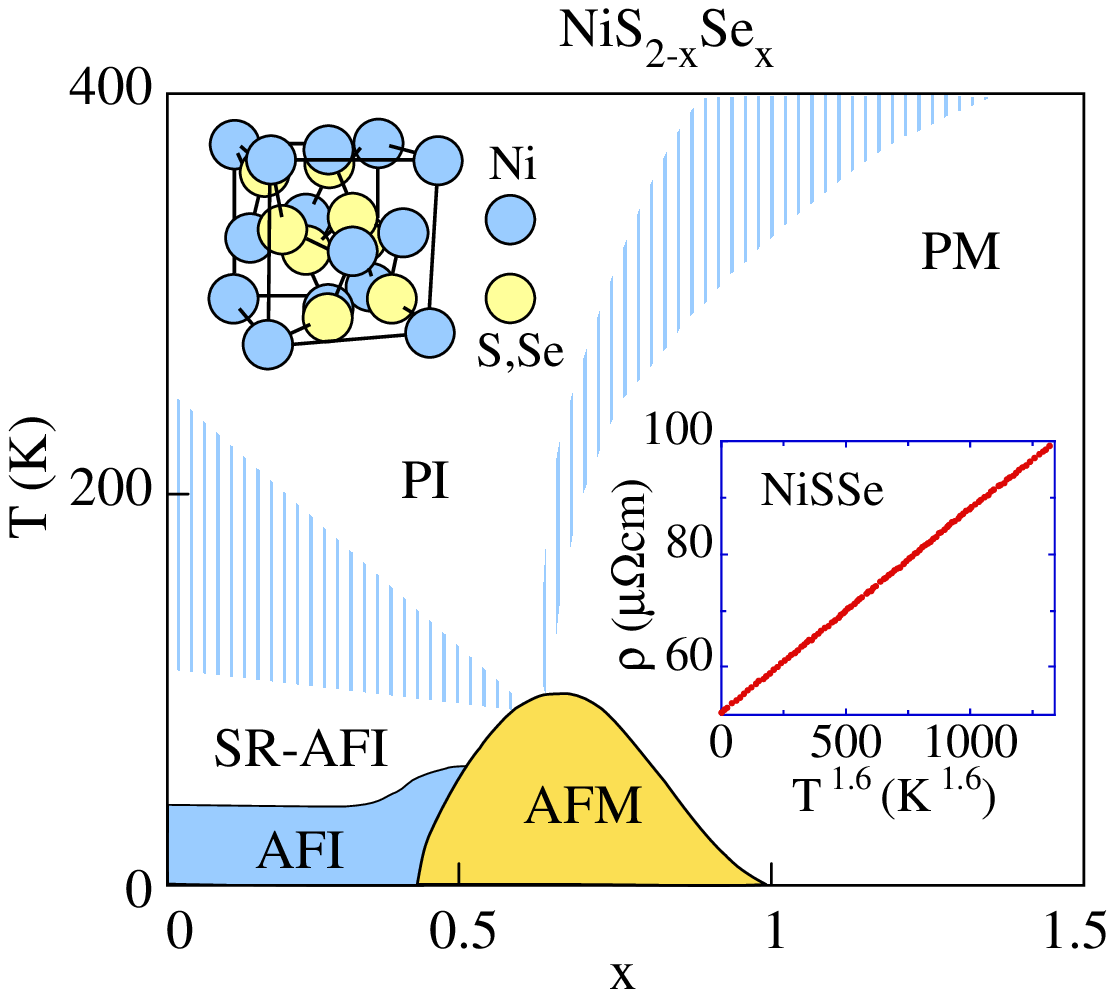}{Temperature-composition phase diagram of
NiS$_{2-x}$Se$_x$.\cite{mat00a,hon98a}\ Lower inset shows the
temperature dependence of the resistivity of NiSSe ($x=1$), which
is just on the border of a metallic antiferromagnetic state at low
temperatures.\cite{miy00a}\ The upper inset is the pyrite crystal
structure of NiS$_2$ and NiSe$_2$.  PI and PM stand for
paramagnetic insulator and paramagnetic metal, respectively. AFI
and AFM stand for antiferromagnetic insulator and
antiferromagnetic metal, respectively.  SR-AFI stands for short
range antiferromagnetic insulator.\cite{hon98a}\ The
temperature-pressure phase diagram of NiS$_2$ is expected to be
similar in form to the temperature-composition phase diagram of
NiS$_{2-x}$Se$_x$. In this study, the AFM boundary of NiS$_2$ is
found to be at $p_c=76\pm 5$~kbar. Thus, NiS$_2$ at $p_c$ and
NiSSe at ambient pressure may be expected to be similar except for
the level of quenched disorder, which is normally higher in NiSSe
than in NiS$_2$. The residual resistivity of NiSSe is typically an
order of magnitude higher than in NiS$_2$.}

In practice, this form of $\tau_{tr}^{-1}$ or of the resistivity
$\rho(T)$ can be restricted to a relatively narrow range of
temperatures and pressures where $\kappa(T)$ is small compared
with the characteristic wavevector of thermally excited magnetic
fluctuations.  The SCR model is not restricted to this limit and
can in principle include the effects of $\kappa(T)$ and even
determine $\kappa(T)$ in a self-consistent fashion. We note that
the SCR model assumes implicitly that an effective underlying
mechanism exists to remove momentum from the electron system. The
validity of this assumption has not been clearly confirmed
theoretically, but seems to be consistent with experiment in the
cases mentioned above. In its conventional form the model does not
include the possible effects of inhomogeneities or texture that
may arise on the border of first order ferromagnetic transitions
as in, e.g., MnSi.\cite{pfl01b,doi03b,bel99a} The SCR model is
also expected to fail on the border of electron localization, as
near a Mott transition or close to local quantum critical points
in heavy electron compounds.\cite{col01a,siq01a}

The applicability of the SCR model has also been questioned for
the case of itinerant-electron antiferromagnetism in general, even
well away from the border of a Mott
transition.\cite{hlu95a,ros99a,ros00a}\ In this paper we present
an attempt to test the prediction of the SCR model in a metal on
the border of metallic antiferromagnetism in 3D for which $\rho$
is predicted to vary as $T^{3/2}$ in the idealized limit
$\kappa\rightarrow 0$, $T\rightarrow 0$, where $\kappa$ now stands
for the correlation wavevector for the staggered magnetization.
The exponent of $3/2$ is the ratio of the spatial dimension $d=3$
and the dynamical exponent $z=2$. This may be contrasted with the
corresponding exponent of $5/3$, which is the ratio of $(d+2)=5$
and $z=3$, for the border of metallic ferromagnetism. (The 2 in
$d+2$ arises from the effect of small-angle scattering that is
absent in the case of the staggered magnetization.) This simple
model for the scattering from antiferromagnetic fluctuations
assumes that the scattering rate can be averaged over the Fermi
surface. Within the SCR model this procedure would seem to require
the presence of a sufficient level of quenched disorder assumed to
have only the simple consequence of inhibiting the short
circuiting caused by the carrier on the cold spots of the Fermi
surface, i.e., regions far from the hot spots connected by the
antiferromagnetic ordering wavevector and thus strongly affected
by spin-fluctuation scattering.\cite{hlu95a,ros99a,ros00a}\

The effect of quenched disorder on the temperature dependence of
$\rho$ in the SCR model shows up particularly clearly in the
temperature-dependent resistivity exponent defined as
$n=\partial\ln\Delta\rho/\partial\ln T$, where
$\Delta\rho=\rho(T)-\rho_0$ and $\rho_0$ is the residual
resistivity, i.e., the resistivity extrapolated to $T=0$~K. The
resistivity exponent $n$ may be described in terms of the Fermi
wavevector $k_F$, the elastic mean free path $l$ of charge
carriers and the reduced temperature $t=T/T_{sf}$, where $T_{sf}$
is a characteristic spin fluctuation temperature.\cite{mat68a}\
With increasing $t$, $n$ drops from $3/2$ towards unity around
$t\approx 1/(k_Fl)$, back towards a value of the order of $3/2$
around $t\approx 1/\sqrt{k_Fl}$ and then towards zero for $t>>
1/(k_Fl)$.\cite{hlu95a,ros99a,ros00a}\ This undulating behaviour
of $n$ is a dramatic prediction of the model for relatively pure
samples that could in principle be tested by studying a series of
samples of the same material with different levels of quenched
disorder. Rosch\cite{ros99a,ros00a}\ introduced this model in an
effort to understand the temperature dependence of the resistivity
exponent measured for the f-electron metal
CePd$_2$Si$_2$.\cite{jul98a,gro00a}\ The results were, in this
case, not entirely conclusive because of the anisotropic and
complex nature of the spin fluctuation spectrum in this material.

Here we present a test of Rosch's model in a d-electron system
with a cubic structure and $T_{sf}$ one to two orders of magnitude
greater than in typical heavy f-electron systems. We compare the
temperature dependence of $\rho$ of two closely related systems on
the border of antiferromagnetism. The two materials are NiS$_2$
near 76~kbar and NiS$_{2-x}$Se$_x$ for $x=1$ where the N\'eel
temperature $T_N$ vanishes at ambient pressure. Due to random
variations in the locations of the S and Se atoms in the lattice,
values of $\rho_0$ found in NiSSe are typically one order of
magnitude higher than in stoichiometric NiS$_2$ compounds.

NiS$_2$ crystallizes in the cubic pyrite structure and is an
antiferromagnetic insulator at ambient pressure at low
temperatures.\cite{bul82a,mat98b}\ It can be metallized via the
application of pressure\cite{wil71a,wil72a,wil85a,sek97a}\ or by
Se
substitution.\cite{wil71a,wil72a,wil85a,mat00a,miy00a,mor83a,hus00a}\
The temperature-pressure phase diagram of NiS$_2$ and the
temperature-composition phase diagram of NiS$_{2-x}$Se$_x$
(Fig.~\ref{ppnissephdiag}) are expected to be similar. Of
particular interest for this paper is the boundary of metallic
antiferromagnetism that appears (i) at $x\cong 1$, i.e., for
NiSSe, in temperature-composition phase diagram
(Fig.~\ref{ppnissephdiag}) and (ii) at $p_c$ in the
temperature-pressure phase diagram for stoichiometric NiS$_2$. We
note that in both cases the quantum critical point for metallic
antiferromagnetism arises well beyond the metal insulator
transition (see Fig.~\ref{ppnissephdiag}\ for the case where
composition is the control parameter).

We present a high-pressure study of the temperature dependence of
the resistivity of NiS$_2$ in the metallic state, which reveals a
critical pressure  $p_c\cong 76$~kbar. The results at $p_c$ are
compared with that of NiSSe at ambient pressure reported in
Ref.~\onlinecite{miy00a}. In both materials one observes a
non-Fermi liquid form of the resistivity that appears in first
approximation to be consistent with the predictions of the SRC
model (inset of Fig.~\ref{ppnissephdiag}\ for NiSSe). However, the
resistivity of NiS$_2$ near $p_c$ has an undulating component in
its variation with temperature that is not seen in NiSSe. This
difference in behaviour might arise from the effects of cold and
hot spots on the Fermi surface as suggested by the model due to
Rosch discussed above.\cite{ros99a,ros00a}\

\section{Experimental}

Single crystals of NiS$_2$ have been grown by chemical vapour
transport using iodine as the transport agent.\cite{wil71a}\ The
residual resistivity ratio $(\rho(273\mbox{~K})/\rho_0)$ above
50~kbar is about 30 for our samples, compared with about 3 for
NiSSe.\cite{miy00a}\ The carrier mean free path of NiS$_2$ is thus
expected to be about an order of magnitude higher than in NiSSe.

Pressure was applied by means of a non-magnetic Bridgman cell
using tungsten carbide anvils. The cell used is a scaled down
version of that designed by Wittig.\cite{wit66a}\ The cullet
diameter of the anvils was 3.5~mm. The gasket was made of
pyrophyllite with a central hole to accommodate the NiS$_2$ sample
as well as a Pb sample and the steatite pressure-transmission
medium. After compression, the sample space was reduced to about
1.5~mm in diameter and 0.1~mm in thickness. The pressure was
determined from the superconducting transition temperature of the
Pb sample.\cite{eic68a}\ The width of the superconducting
transition suggested that the pressure variation over the Pb
sample was about 10\%\ of the average pressure.

The resistivity was measured via the 4-terminal ac technique. Four
50~\micron\ Pt leads were passed into the high-pressure region via
grooves in the insulating pyrophyllite gasket.  The bare wires
were rested on top of the samples and pressed onto the sample
surface during pressurization to achieve adequate electrical
contacts.

The resistivity measurements were carried out with two different
and independent systems, a helium circulation cryostat (ILL Orange
cryostat; 1.5~K to 300~K) in Grenoble and an adiabatic
demagnetization refrigerator (0.04~K to 100~K) in Cambridge. The
latter system had two voltage channels, one with a low temperature
transformer and the other with a room-temperature transformer. The
excitation currents were 1~mA and 160~$\mu$A in the Orange
cryostat and in the adiabatic demagnetization fridge,
respectively. The results obtained with these two experimental
systems were consistent with each other where comparisons could be
made.

\section{Results}

\pgnfigure{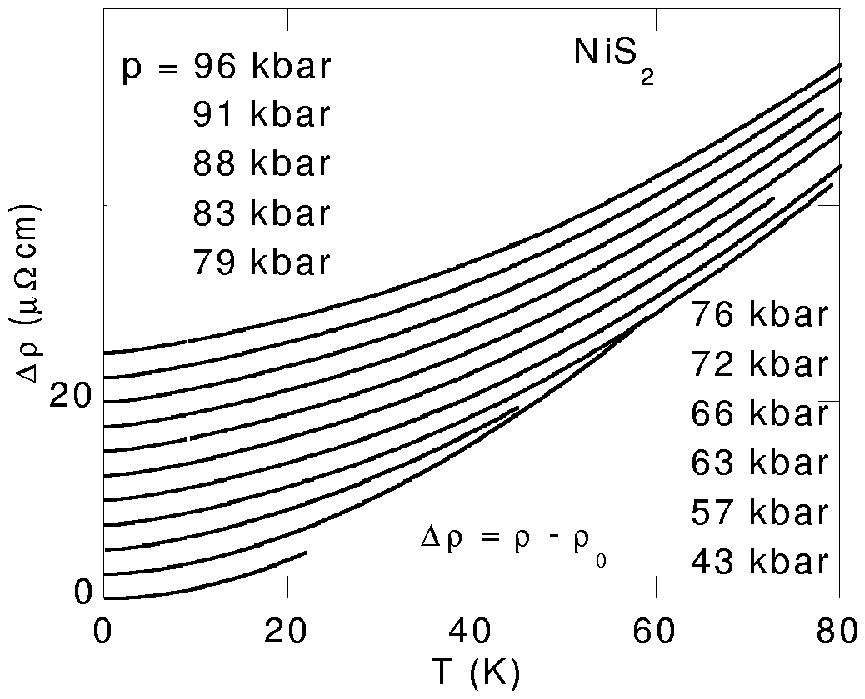}{Temperature dependence of the resistivity in
the metallic state of NiS$_2$ at high pressures. $\Delta\rho$ is
$\rho(T)-\rho_0$, where $\rho_0$ is the residual resistivity
extrapolated to $T=0$~K. The curves are shifted vertically for
clarity.}

\pgnfigurestwo{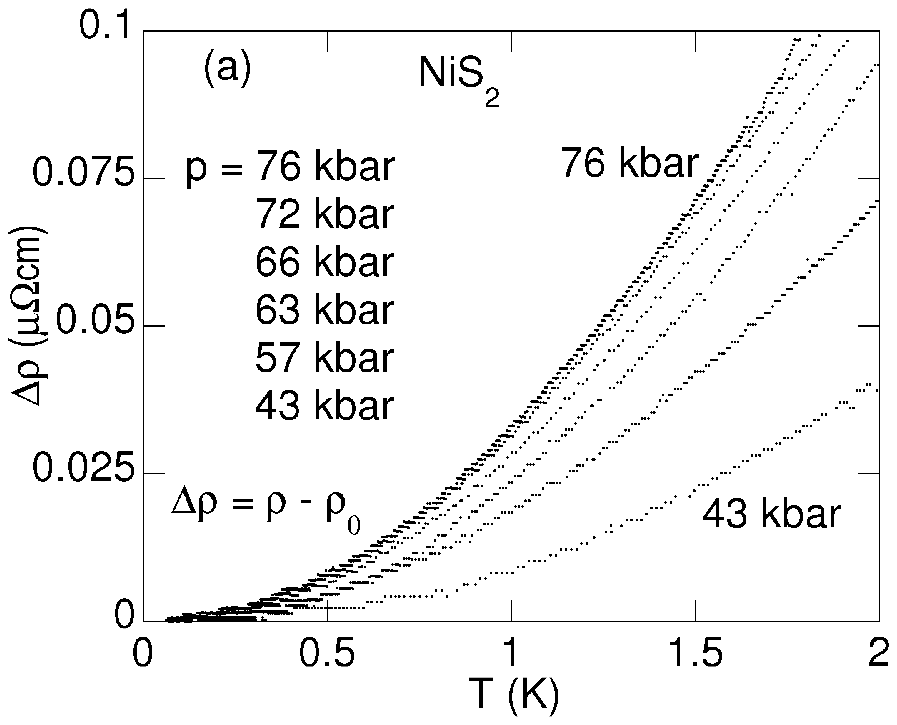}{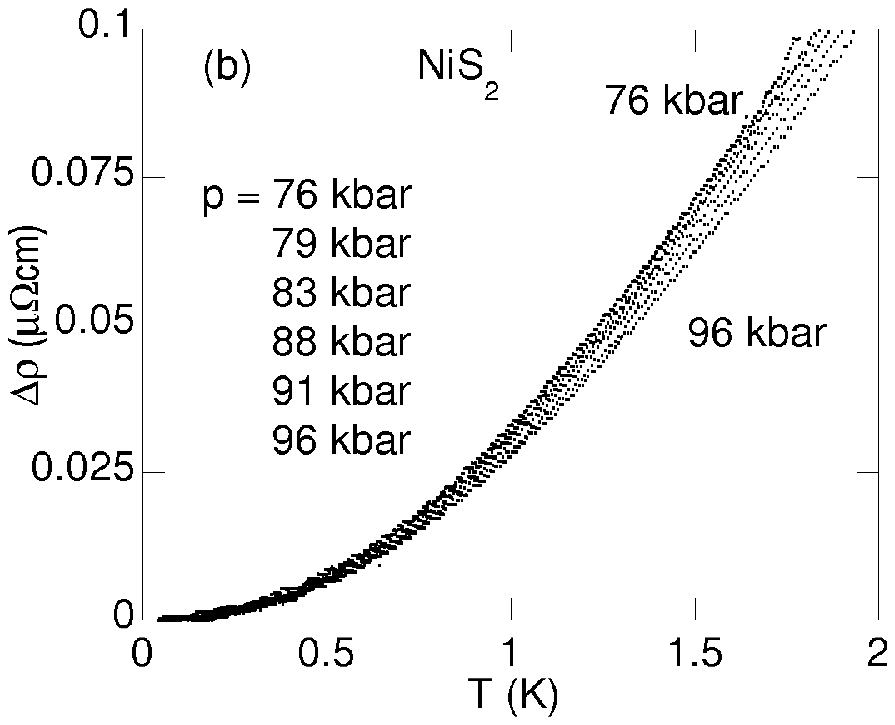}{Low temperature
variation of the resistivity in the metallic state of NiS$_2$ at
high pressures. $\Delta\rho$ grows in strength up to 76~kbar (a)
and weakens gradually above 76~kbar (b).}

\pgnfigure{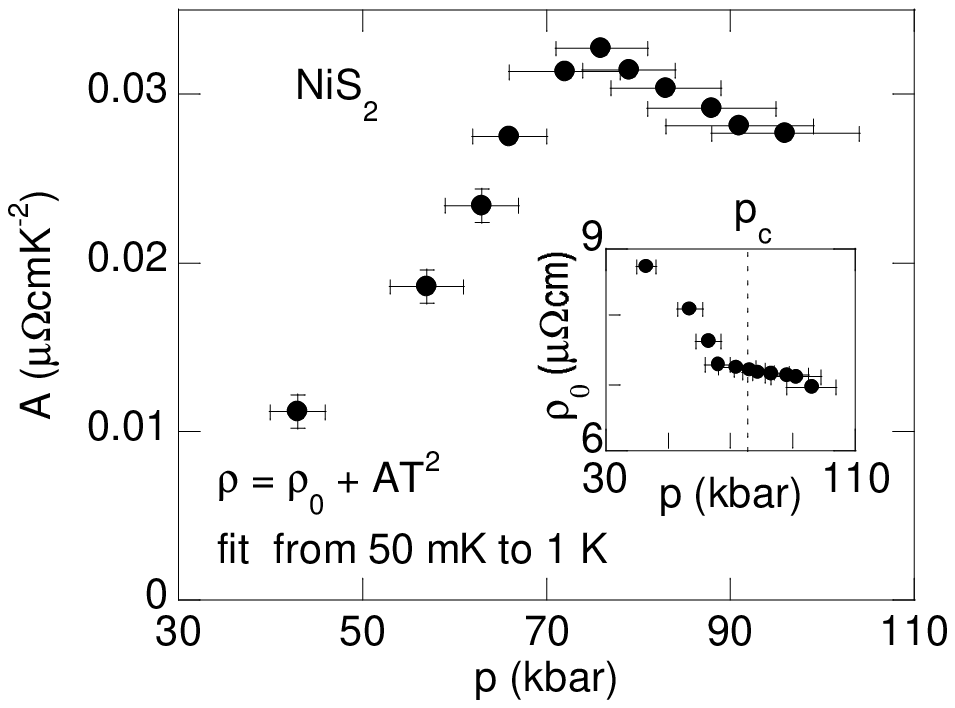}{The $T^2$ coefficient of the
resistivity, $A$, and residual resistivity, $\rho_0$ (inset), of
NiS$_2$ in the metallic high-pressure state. The parameters $A$
and $\rho_0$ are obtained by a fit of the resistivity as indicated
in the figure. $A$ is peaked at $p_c=76\pm 5$~kbar. This marks the
boundary of the metallic antiferromagnetic state of NiS$_2$.}

\pgnfiguresthree{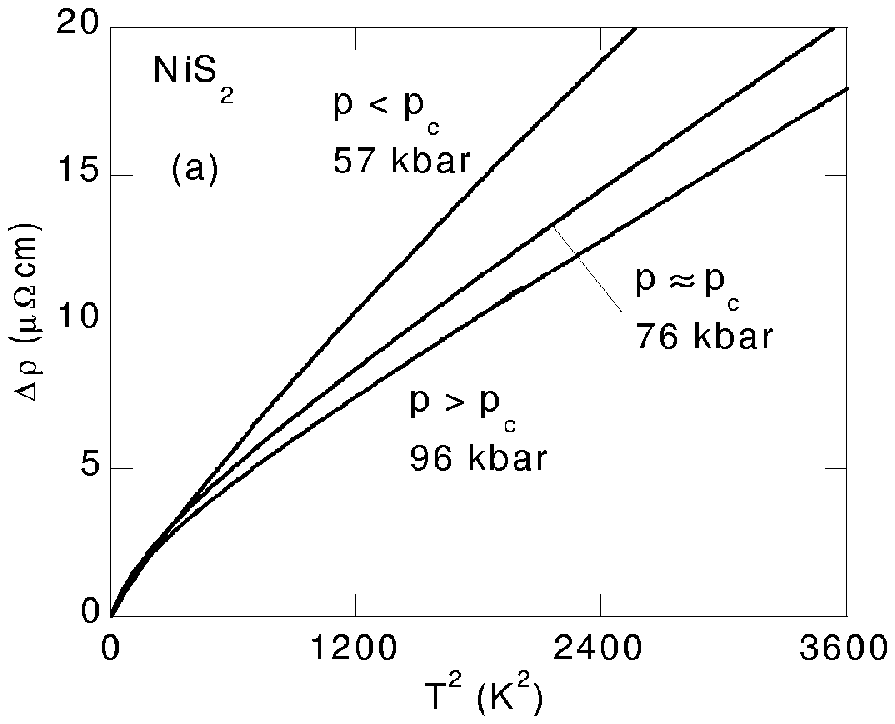}{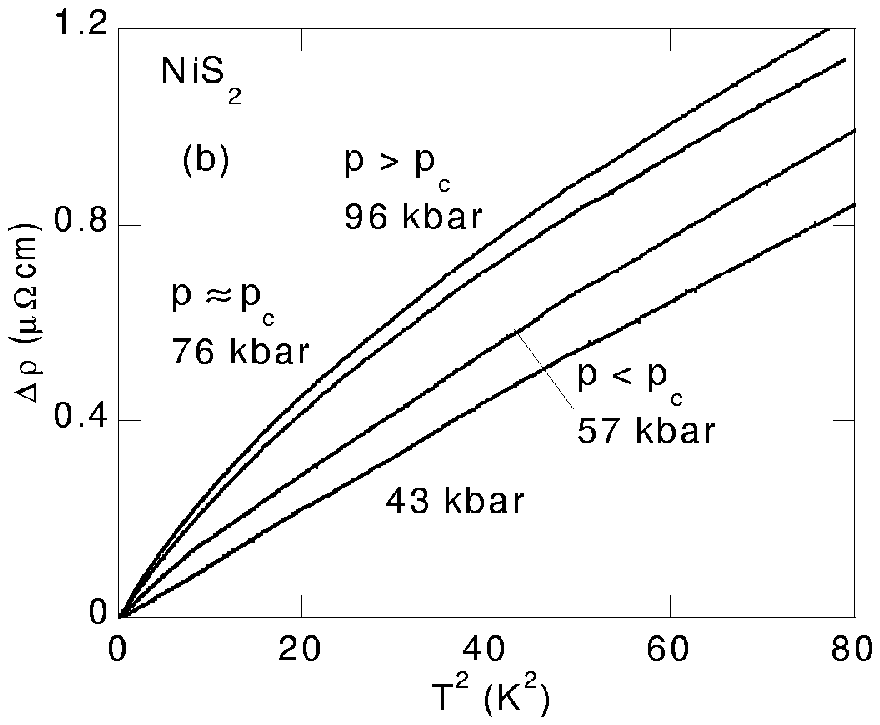}{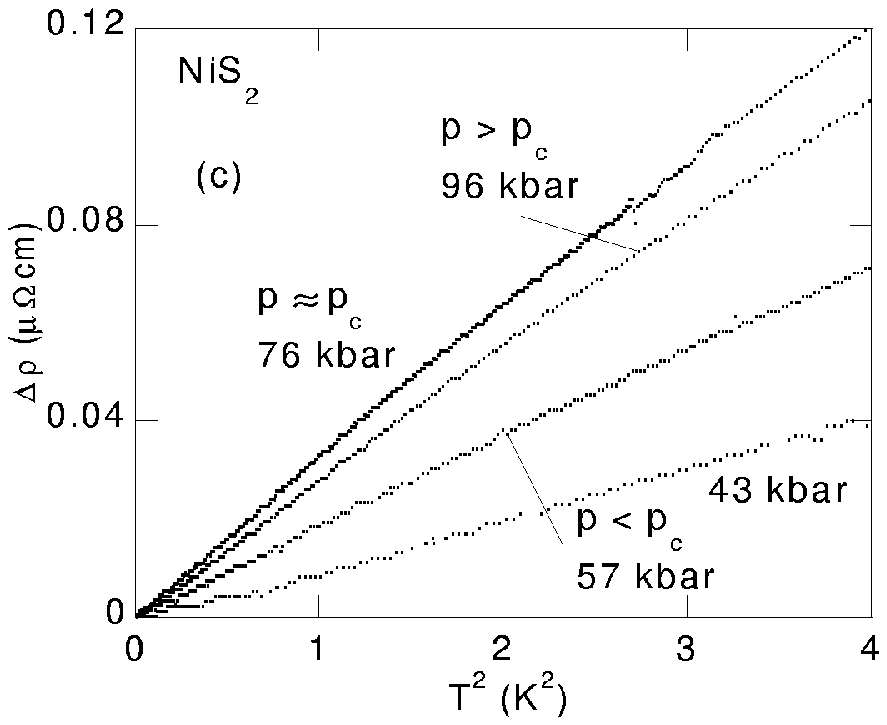}{Resistivity
vs $T^2$ in the high pressure metallic regime of NiS$_2$. The
three panels are for different temperature intervals. A quadratic
temperature dependence of $\Delta\rho$ is seen only at the lowest
temperatures (below a few K) and the $T^2$ coefficient of
$\Delta\rho$ is peaked at $p_c\approx 76$~kbar as shown in
Fig.~\ref{ppnisssqcoeff}.}

\pgnfiguresthree{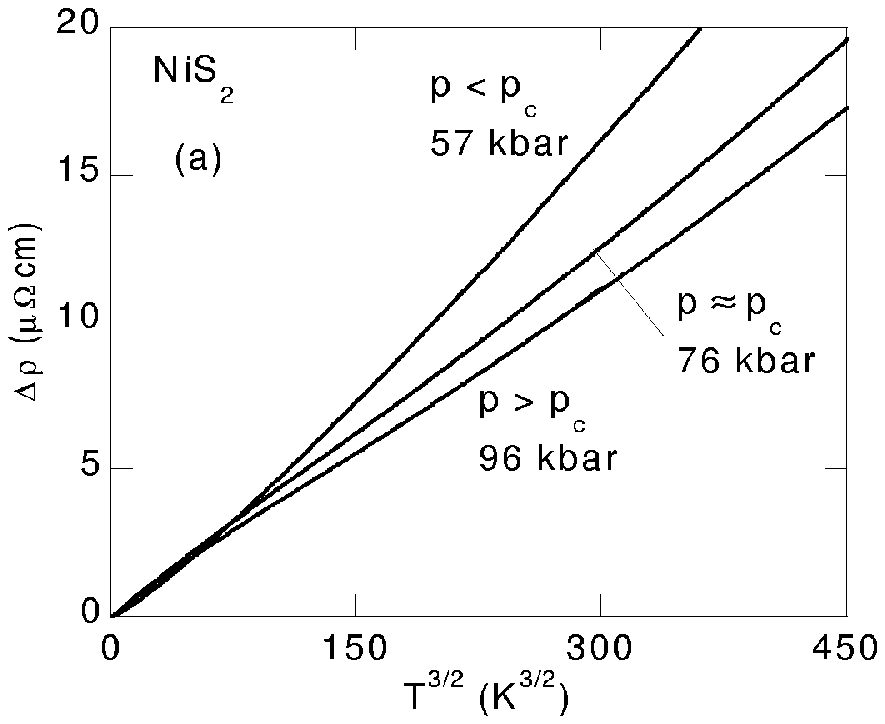}{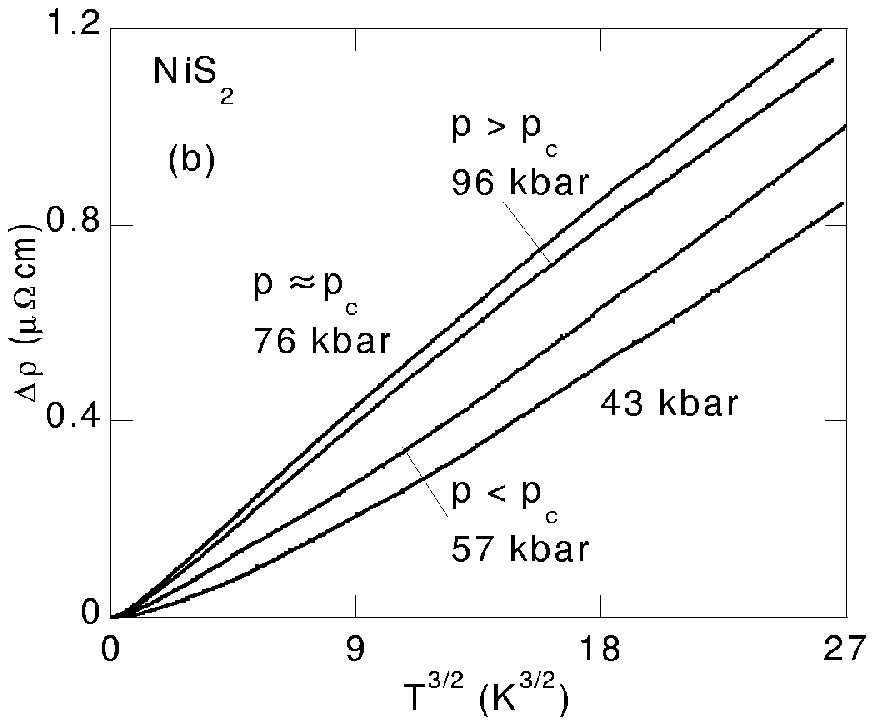}{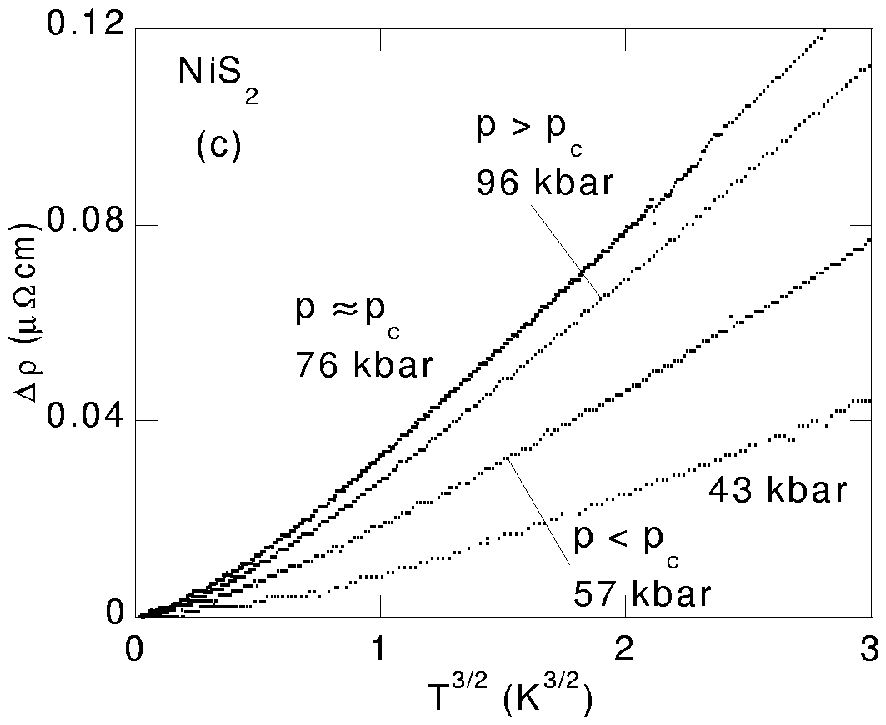}{Resistivity
vs $T^{3/2}$ in the high-pressure metallic regime of NiS$_2$. The
three panels are for different temperature intervals. An
approximately $T^{3/2}$ variation of $\Delta\rho$ is seen around
$p_c\approx 76$~kbar only over a decade in temperature above a few
K. $\Delta\rho$ is quadratic in temperature below a few K at all
pressures in the metallic regime studied (see
Fig.~\ref{ppnisssqht}c).}

The temperature variations of the resistivity of NiS$_2$ in the
high pressure metallic state above 40~kbar are presented in
Figures~\ref{ppnissov}\ and \ref{ppnissovltlp}. (The
insulator-to-metal transition (not shown) was observed at around
30~kbar, in agreement with the literature.\cite{wil71a})
Figure~\ref{ppnissov}\ shows $\Delta\rho$ vs $T$ up to 80~K and
Figure~\ref{ppnissovltlp}\ is an expanded view of $\Delta\rho$ vs
$T$ in the range 0.05~K to 2~K. Plots of $\Delta\rho$ vs $T^2$ and
$\Delta\rho$ vs $T^{3/2}$ over different ranges in temperatures
are shown, respectively, in Figures~\ref{ppnisssqht}\ and
\ref{ppnissthht}.

Below 1~K the resistivity can be described by an equation of the
form $\rho=\rho_0+AT^2$ over the entire pressure range explored,
43~kbar to 96~kbar. The pressure variations of the fitted values
of $A$ and $\rho_0$ are given in Figure~\ref{ppnisssqcoeff}. The
$T^2$ coefficient $A$ exhibits a peak at $p_c=76\pm 5$~kbar.

\pgnfigure{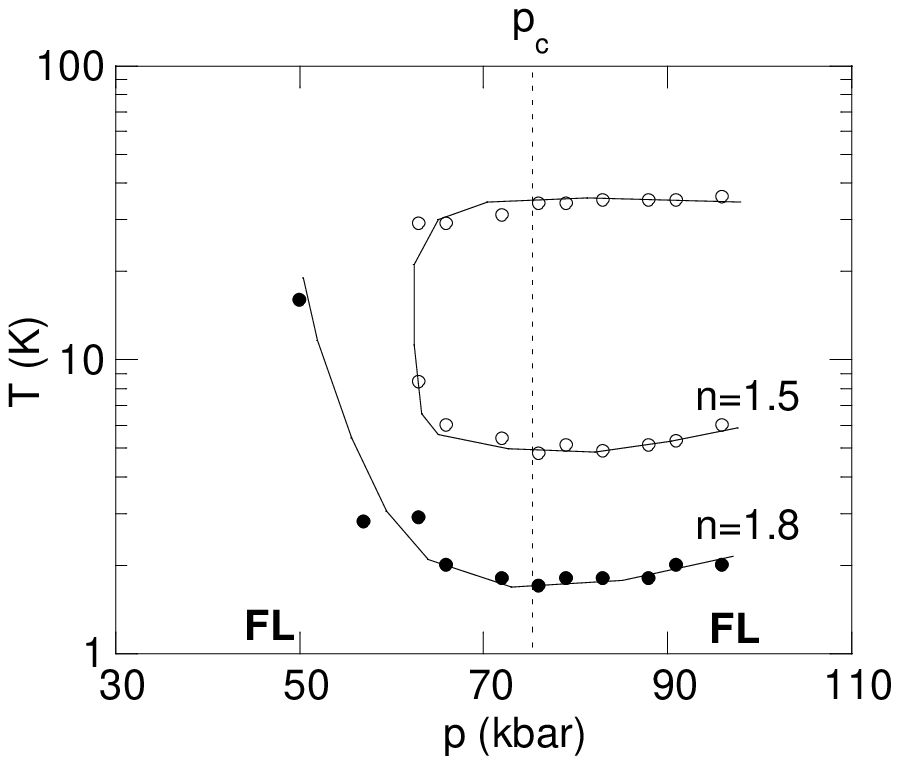}{Contours of the resistivity exponent
$n=\partial\ln\Delta\rho/\partial\ln T$ in the
temperature-pressure plane for NiS$_2$ in the metallic regime. A
Fermi liquid (FL) $T^2$ temperature dependence of the resistivity
is seen at the lowest temperatures up to $T_{FL}$ defined,
somewhat arbitrarily, by the condition $n(T_{FL})=1.8$. $T_{FL}$
reaches a minimum at $p_c\approx 76$~kbar. A non-Fermi liquid
$T^n$ form of the resistivity with $n\approx 3/2$ is seen over
approximately a decade in temperature above a few K near $p_c$.}

\pgnfigurestwo{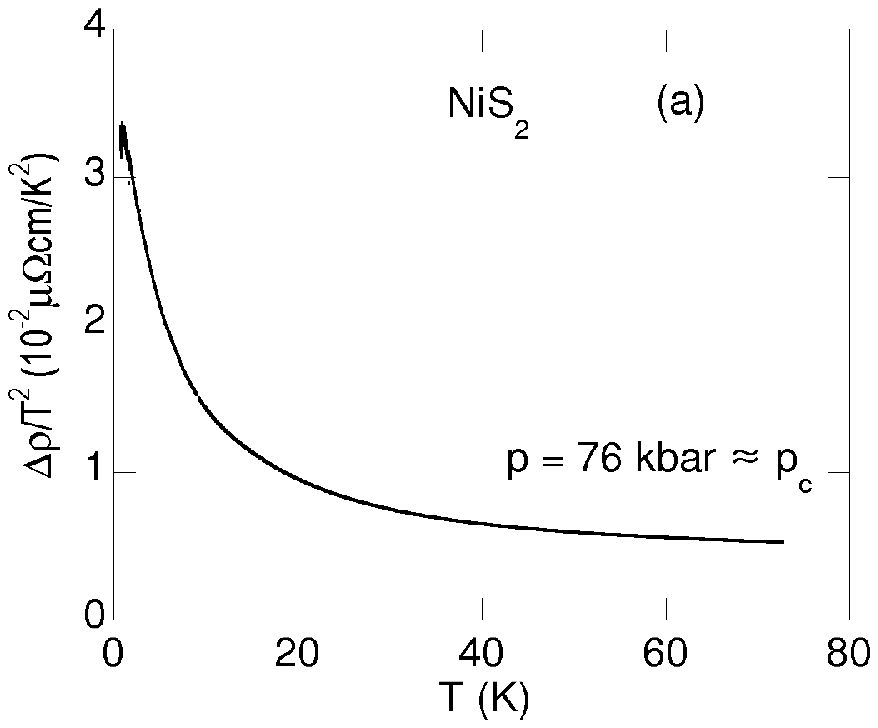}{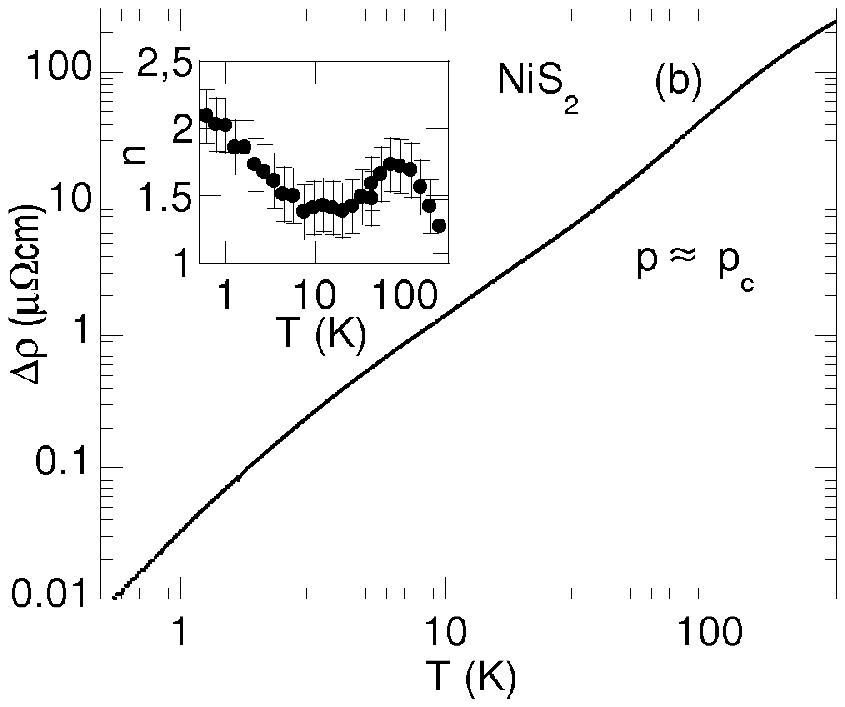}{The non-Fermi liquid
form of the resistivity of NiS$_2$ near $p_c\approx 76$~kbar. (a)
$\Delta\rho/T^2$ vs $T$ does not saturate and thus has a non-Fermi
liquid form over a wide temperature range, except at very low
temperatures (Fig.~\ref{ppnisssqht}). (b) A plot of
$\ln\Delta\rho$ vs $\ln T$ and the resistivity exponent
$n=\partial\ln\Delta\rho/\partial\ln T)$ vs $\ln T$ exhibits an
undulating form in contrast to the simple $T^{3/2}$ temperature
dependence seen in NiSSe (Ref.~\onlinecite{miy00a}\ and
Fig.~\ref{ppnissephdiag}). This undulating form of the resistivity
has been seen in different NiS$_2$ specimens and with two
independent measurement systems.}

Figures~\ref{ppnisssqht}\ and \ref{ppnissthht}\ compare
$\Delta\rho$ vs $T^2$ and $\Delta\rho$ vs $T^{3/2}$, respectively,
in three panels each covering different ranges in temperature.
Figure~\ref{ppnisssqht}c, in particular, highlights the $T^2$
variation observed at all pressures with a peak of $A$ at around
$p_c$ as discussed above. Figures~\ref{ppnissthht}a,
\ref{ppnissthht}b\ and \ref{ppnissncontour}, on the other hand,
suggest that near this pressure $\Delta\rho$ varies roughly as
$T^{3/2}$ over a decade in temperature above a few K. This is the
behaviour predicted by the SCR model for a system on the border of
metallic antiferromagnetism as discussed in the introduction. The
existence of a $T^2$ regime even at $p_c$, however, suggests that
the transition into the antiferromagnetic state below $p_c$ may
not be continuous, i.e., that the antiferromagnetic quantum
critical point is not quite reached due to the onset of a first
order transition. We note that the weak pressure variation of the
Fermi liquid crossover temperature $T_{FL}$ near $p_c$ (see
Fig.~\ref{ppnissncontour}\ and the caption) would seem to rule out
an explanation of the $T^2$ resistivity in terms of an
inhomogeneity in pressure.

The identification of $p_c$ with the boundary of metallic
antiferromagnetism is also suggested by the correspondence of the
temperature-pressure and temperature-composition phase diagram of
NiS$_2$ found in earlier work.\cite{mat00a,miy00a,mor83a,hus00a}\
This predicted that the critical pressure for the border of
metallic antiferromagnetism should be around 60~kbar, which is of
the same order of magnitude as $p_c$ defined above. The
identification of $p_c$ with the antiferromagnetic boundary in the
metallic state is tentative and needs confirmation by other
measurements and in particular the detection of a signature of
$T_N$ in the resistivity for pressures below $p_c$. Here we focus
attention mainly on the behaviour of $\rho$ near $p_c$ and
contrast it with that of NiSSe at ambient pressure (inset of
Fig.~\ref{ppnissephdiag}).

The non-Fermi liquid behaviour of $\Delta\rho$ over a wide
temperature range near $p_c$ (low-temperature data at 76~kbar with
added 77~kbar data up to 300~K; the high-temperature data is
scaled to match the low-temperature data) is highlighted in
Figure~\ref{ppnisspcsqdev}.  Panel (a) shows the dramatic upturn
of $\Delta\rho$/$T^2$, which in the Fermi liquid regime is
expected to saturate to a constant value. Panel (b) is a plot of
$\ln\Delta\rho$ vs $\ln T$ which shows that the average slope
corresponds to a resistivity exponent close to $3/2$, as discussed
above and as seen in NiSSe at ambient pressure. However, in
contrast to NiSSe, the temperature dependence of the resistivity
in NiS$_2$ at $p_c$ exhibits an undulating structure which is
evident in Figure~\ref{ppnisspcsqdev}\ and highlighted in the
temperature dependence of the resistivity exponent
$n=\partial\ln\Delta\rho/\partial\ln T$ shown in the inset. We
note that this undulating structure has been observed in several
samples of NiS$_2$ and in two independent measurement systems.

\section{Discussion}

The resistivity measurements suggest that antiferromagnetism in
the high-pressure metallic state of NiS$_2$ is suppressed at a
critical prssure of $p_c=76\pm 5$~kbar. At $p_c$, we find that the
$T^2$ coefficient of the resistivity $A$ has its maximum and the
Fermi liquid crossover temperature $T_{FL}$ defined in the caption
of Figure~\ref{ppnissncontour}\ has its minimum. Above $T_{FL}$
near $p_c$ the resistivity has an approximately $T^{3/2}$
temperature dependence over a decade in temperature.  As already
stated, this is the behaviour expected in the SCR model for a
metal in 3D on the border of antiferromagnetism at low
temperatures. Also, we note that $p_c$ is of the same order of
magnitude as that inferred for the correspondence between the
temperature-composition and temperature-pressure phase diagram as
discussed in the previous section.

The weakness or absence of a signature of $T_N$ in the resistivity
below $p_c$ is not necessarily surprising since the same is the
case for NiS$_{2-x}$Se$_x$ in the range $0.4\le x\le 1$
(Refs.~\onlinecite{yao96a,miy00a}) where antiferromagnetic order
has been confirmed by neutron
scattering.\cite{sud85a,sud92a,mat00a}\ The signature of $T_N$ in
$\rho$ can be washed out by inhomogeneities in composition or
pressure and may only show up in high-precision measurements of
$\partial\rho/\partial T$ in the pressure range where $T_N$ is not
too strongly varying with pressure and where pressure is
hydrostatic.

The fact that $T_{FL}$ and correspondingly the $T^2$ coefficient
of the resistivity $A$ remain finite at $p_c$ is not necessarily
inconsistent with our assumption that $p_c$ marks the border of
antiferromagnetism. The antiferromagnetic transition vs pressure
may be first order as is found in a number of antiferromagnetic
metals such as YMn$_2$ and GdMn$_2$.\cite{hau95a}\ First order
quantum phase transitions also seem to be common in ferromagnetic
metals such as MnSi and Ni$_3$Al.\cite{pfl97a,the97a,nik04d}\
$T_{FL}$ and $A$ remain finite at the antiferromagnetic boundary
of YMn$_2$, a system that shares some features in common with
NiS$_2$. (We note, however, that the absence of Fermi liquid
behaviour does not necessarily mean that the quantum phase
transition is continuous.\cite{pfl01b, doi03b})

The Fermi liquid crossover temperature, $T_{FL}$, in NiS$_2$ is
very much lower than the characteristic spin fluctuation
temperature, $T_{sf}$, which may be expected to be of the order of
$10^3$~K in a typical d-metal. We consider to what extent the
behaviour above $T_{FL}$, but below $T_{sf}$, may be understood in
terms of an itinerant-electron model for a continuous
antiferromagnetic quantum critical point. The predictions of the
SCR model as analyzed by Rosch has already been outlined in the
introduction for an antiferromagnetic quantum critical point. With
increasing reduced temperature $t=T/T_{sf}$, the SCR model
predicts that the resistivity exponent
$n=\partial\ln\Delta\rho/\partial\ln T$ drops from an initial
value of 3/2 towards unity around $t\approx 1/(k_Fl)$, back to a
value of order 3/2 around $t\approx 1/\sqrt{k_Fl}$ and then to
zero for $t>>1/(k_Fl)$.\cite{hlu95a,ros99a,ros00a}\ This behaviour
is qualitatively similar to that seen in NiS$_2$ at $p_c$ and
above $T_{FL}$ (inset Fig.~\ref{ppnisspcsqdev}). For reasonable
choices of parameters for NiS$_2$, $T_{sf}\approx 10^3$~K,
$k_F\approx 0.5$~\AA$^{-1}$ and $l\approx
140$~\AA$^{-1}$,\cite{nik06a}\ we expect the minimum of $n$ to
occur near 15~K and the maximum at around 120~K. These crossover
temperatures are in rough agreement with our observations.

In NiSSe, $l$ is an order of magnitude smaller than in NiS$_2$ and
thus the minimum and maximum of $n$ are expected to arise at
around 100~K and 320~K, respectively. Over the temperature range
of the experiments shown in the inset of
Figure~\ref{ppnissephdiag}\ the model predicts a simple $T^{3/2}$
temperature dependence without modulation, as is seen. The
difference in the behaviour of NiS$_2$ and NiSSe at their
respective critical conditions is thus not surprising. The
combined effects of scattering from spin and lattice fluctuations
might also lead to an undulating behaviour of $n$, but this would
naively be expected to arise in both NiS$_2$ and NiSSe, in
contradiction with observation.

\section{Conclusions}

The temperature dependence of the resistivity of NiS$_2$ near the
critical pressure, which has been found to be $p_c=76\pm 5$~kbar,
is consistent with that expected for a metal on the border of
itinerant-electron antiferromagnetism. The Fermi liquid crossover
temperature, $T_{FL}$, which defines the range over which the
resistivity is roughly quadratic in temperature, does not vanish
at $p_c$, but is three orders of magnitude smaller than the
characteristic spin fluctuation temperature $T_{sf}$. The finite
value of $T_{FL}$ may indicate that the antiferromagnetic quantum
critical point is first order, as is the case in related materials
such as YMn$_2$.

Over a wide temperature range above $T_{FL}$ the temperature
variation of the resistivity exhibits a non-Fermi liquid form that
can be understood in terms of the effects of spin-fluctuation
scattering at cold and hot spots of the Fermi surface as
anticipated by Rosch in his refined treatment of the
self-consistent-renormalization (SCR) model. In particular, the
resistivity exponent $n=\partial\ln\Delta\rho/\partial ln T$
exhibits an undulating structure which is consistent not only
qualitatively, but also approximately quantitatively with the
predictions of this model. The model also accounts for the absence
of this undulating structure in the related, but less pure
material, NiSSe, which is known to be at the border of
antiferromagnetism at ambient pressure.


\begin{acknowledgments}

We thank A. Rosch and S. Julian for valuable discussions. PGN is
grateful for support by the FERLIN program of the European Science
Foundation.

\end{acknowledgments}


\begin{thebibliography}{39}
\expandafter\ifx\csname
natexlab\endcsname\relax\def\natexlab#1{#1}\fi
\expandafter\ifx\csname bibnamefont\endcsname\relax
  \def\bibnamefont#1{#1}\fi
\expandafter\ifx\csname bibfnamefont\endcsname\relax
  \def\bibfnamefont#1{#1}\fi
\expandafter\ifx\csname citenamefont\endcsname\relax
  \def\citenamefont#1{#1}\fi
\expandafter\ifx\csname url\endcsname\relax
  \def\url#1{\texttt{#1}}\fi
\expandafter\ifx\csname
urlprefix\endcsname\relax\def\urlprefix{URL }\fi
\providecommand{\bibinfo}[2]{#2}
\providecommand{\eprint}[2][]{\url{#2}}

\bibitem[{\citenamefont{Hertz}(1976)}]{her76a}
\bibinfo{author}{\bibfnamefont{J.~A.} \bibnamefont{Hertz}},
  \bibinfo{journal}{Phys.Rev.B} \textbf{\bibinfo{volume}{14}},
  \bibinfo{pages}{1165} (\bibinfo{year}{1976}), \bibinfo{note}{and references
  therein}.

\bibitem[{\citenamefont{Millis}(1993)}]{mil93a}
\bibinfo{author}{\bibfnamefont{A.~J.} \bibnamefont{Millis}},
  \bibinfo{journal}{Phys.Rev.B} \textbf{\bibinfo{volume}{48}},
  \bibinfo{pages}{7183} (\bibinfo{year}{1993}), \bibinfo{note}{and references
  therein}.

\bibitem[{\citenamefont{Moriya}(1985)}]{mor85a}
\bibinfo{author}{\bibfnamefont{T.}~\bibnamefont{Moriya}},
  \emph{\bibinfo{title}{Spin Fluctuations in Itinerant Electron Magnetism}}
  (\bibinfo{publisher}{Springer}, \bibinfo{address}{Berlin},
  \bibinfo{year}{1985}), \bibinfo{note}{and references therein}.

\bibitem[{\citenamefont{Lonzarich}(1997)}]{lon97a}
\bibinfo{author}{\bibfnamefont{G.~G.} \bibnamefont{Lonzarich}}, in
  \emph{\bibinfo{booktitle}{Electron}}, edited by
  \bibinfo{editor}{\bibfnamefont{M.}~\bibnamefont{Springford}}
  (\bibinfo{publisher}{Cambridge University Press},
  \bibinfo{address}{Cambridge}, \bibinfo{year}{1997}),
  chap.~\bibinfo{chapter}{6}, pp. \bibinfo{pages}{109--147}.

\bibitem[{\citenamefont{Niklowitz et~al.}(2005)\citenamefont{Niklowitz,
  Beckers, Lonzarich, Knebel, Salce, Thomasson, Bernhoeft, Braithwaite, and
  Flouquet}}]{nik04d}
\bibinfo{author}{\bibfnamefont{P.~G.} \bibnamefont{Niklowitz}},
  \bibinfo{author}{\bibfnamefont{F.}~\bibnamefont{Beckers}},
  \bibinfo{author}{\bibfnamefont{G.~G.} \bibnamefont{Lonzarich}},
  \bibinfo{author}{\bibfnamefont{G.}~\bibnamefont{Knebel}},
  \bibinfo{author}{\bibfnamefont{B.}~\bibnamefont{Salce}},
  \bibinfo{author}{\bibfnamefont{J.}~\bibnamefont{Thomasson}},
  \bibinfo{author}{\bibfnamefont{N.}~\bibnamefont{Bernhoeft}},
  \bibinfo{author}{\bibfnamefont{D.}~\bibnamefont{Braithwaite}},
  \bibnamefont{and} \bibinfo{author}{\bibfnamefont{J.}~\bibnamefont{Flouquet}},
  \bibinfo{journal}{Phys.Rev.B} \textbf{\bibinfo{volume}{72}},
  \bibinfo{pages}{24424} (\bibinfo{year}{2005}).

\bibitem[{\citenamefont{Mathon}(1968)}]{mat68a}
\bibinfo{author}{\bibfnamefont{J.}~\bibnamefont{Mathon}},
  \bibinfo{journal}{Proc.Roy.Soc.A} \textbf{\bibinfo{volume}{306}},
  \bibinfo{pages}{355} (\bibinfo{year}{1968}).

\bibitem[{\citenamefont{Varma et~al.}(1989)\citenamefont{Varma, Littlewood,
  Schmitt-Rink, Abrahams, and Ruckenstein}}]{var89a}
\bibinfo{author}{\bibfnamefont{C.~M.} \bibnamefont{Varma}},
  \bibinfo{author}{\bibfnamefont{P.~B.} \bibnamefont{Littlewood}},
  \bibinfo{author}{\bibfnamefont{S.}~\bibnamefont{Schmitt-Rink}},
  \bibinfo{author}{\bibfnamefont{E.}~\bibnamefont{Abrahams}}, \bibnamefont{and}
  \bibinfo{author}{\bibfnamefont{A.~E.} \bibnamefont{Ruckenstein}},
  \bibinfo{journal}{Phys.Rev.Lett.} \textbf{\bibinfo{volume}{63}},
  \bibinfo{pages}{1996} (\bibinfo{year}{1989}).

\bibitem[{\citenamefont{Holstein et~al.}(1973)\citenamefont{Holstein, Norton,
  and Pincus}}]{hol73a}
\bibinfo{author}{\bibfnamefont{T.}~\bibnamefont{Holstein}},
  \bibinfo{author}{\bibfnamefont{R.~E.} \bibnamefont{Norton}},
  \bibnamefont{and} \bibinfo{author}{\bibfnamefont{P.}~\bibnamefont{Pincus}},
  \bibinfo{journal}{Phys.Rev.B} \textbf{\bibinfo{volume}{8}},
  \bibinfo{pages}{2647} (\bibinfo{year}{1973}).

\bibitem[{\citenamefont{Baym and Pethick}(1991)}]{bay91a}
\bibinfo{author}{\bibfnamefont{G.}~\bibnamefont{Baym}} \bibnamefont{and}
  \bibinfo{author}{\bibfnamefont{C.}~\bibnamefont{Pethick}},
  \emph{\bibinfo{title}{Landau-Fermi liquid theory}}
  (\bibinfo{publisher}{Wiley}, \bibinfo{address}{New York},
  \bibinfo{year}{1991}), chap.~\bibinfo{chapter}{3}.

\bibitem[{\citenamefont{Matsuura et~al.}(2000)\citenamefont{Matsuura, Hiraka,
  Yamada, and Endoh}}]{mat00a}
\bibinfo{author}{\bibfnamefont{M.}~\bibnamefont{Matsuura}},
  \bibinfo{author}{\bibfnamefont{H.}~\bibnamefont{Hiraka}},
  \bibinfo{author}{\bibfnamefont{K.}~\bibnamefont{Yamada}}, \bibnamefont{and}
  \bibinfo{author}{\bibfnamefont{Y.}~\bibnamefont{Endoh}},
  \bibinfo{journal}{J.Phys.Soc.Jpn.} \textbf{\bibinfo{volume}{69}},
  \bibinfo{pages}{1503} (\bibinfo{year}{2000}).

\bibitem[{\citenamefont{Honig and Spa{\l}ek}(1998)}]{hon98a}
\bibinfo{author}{\bibfnamefont{J.~M.} \bibnamefont{Honig}} \bibnamefont{and}
  \bibinfo{author}{\bibfnamefont{J.}~\bibnamefont{Spa{\l}ek}},
  \bibinfo{journal}{Chem.Mater.} \textbf{\bibinfo{volume}{10}},
  \bibinfo{pages}{2910} (\bibinfo{year}{1998}).

\bibitem[{\citenamefont{Miyasaka et~al.}(2000)\citenamefont{Miyasaka, Takagi,
  Sekine, Takahashi, Mori, and Cava}}]{miy00a}
\bibinfo{author}{\bibfnamefont{S.}~\bibnamefont{Miyasaka}},
  \bibinfo{author}{\bibfnamefont{H.}~\bibnamefont{Takagi}},
  \bibinfo{author}{\bibfnamefont{Y.}~\bibnamefont{Sekine}},
  \bibinfo{author}{\bibfnamefont{H.}~\bibnamefont{Takahashi}},
  \bibinfo{author}{\bibfnamefont{N.}~\bibnamefont{Mori}}, \bibnamefont{and}
  \bibinfo{author}{\bibfnamefont{R.~J.} \bibnamefont{Cava}},
  \bibinfo{journal}{J.Phys.Soc.Jpn.} \textbf{\bibinfo{volume}{69}},
  \bibinfo{pages}{3166} (\bibinfo{year}{2000}).

\bibitem[{\citenamefont{Pfleiderer et~al.}(2001)\citenamefont{Pfleiderer,
  Julian, and Lonzarich}}]{pfl01b}
\bibinfo{author}{\bibfnamefont{C.}~\bibnamefont{Pfleiderer}},
  \bibinfo{author}{\bibfnamefont{S.~R.} \bibnamefont{Julian}},
  \bibnamefont{and} \bibinfo{author}{\bibfnamefont{G.~G.}
  \bibnamefont{Lonzarich}}, \bibinfo{journal}{Nature}
  \textbf{\bibinfo{volume}{414}}, \bibinfo{pages}{427} (\bibinfo{year}{2001}).

\bibitem[{\citenamefont{Doiron-Leyraud
  et~al.}(2003)\citenamefont{Doiron-Leyraud, Walker, Taillefer, Steiner,
  Julian, and Lonzarich}}]{doi03b}
\bibinfo{author}{\bibfnamefont{N.}~\bibnamefont{Doiron-Leyraud}},
  \bibinfo{author}{\bibfnamefont{I.~R.} \bibnamefont{Walker}},
  \bibinfo{author}{\bibfnamefont{L.}~\bibnamefont{Taillefer}},
  \bibinfo{author}{\bibfnamefont{M.~J.} \bibnamefont{Steiner}},
  \bibinfo{author}{\bibfnamefont{S.~R.} \bibnamefont{Julian}},
  \bibnamefont{and} \bibinfo{author}{\bibfnamefont{G.~G.}
  \bibnamefont{Lonzarich}}, \bibinfo{journal}{Nature}
  \textbf{\bibinfo{volume}{425}}, \bibinfo{pages}{595} (\bibinfo{year}{2003}).

\bibitem[{\citenamefont{Belitz et~al.}(1999)\citenamefont{Belitz, Kirkpatrick,
  and Vojta}}]{bel99a}
\bibinfo{author}{\bibfnamefont{D.}~\bibnamefont{Belitz}},
  \bibinfo{author}{\bibfnamefont{T.~R.} \bibnamefont{Kirkpatrick}},
  \bibnamefont{and} \bibinfo{author}{\bibfnamefont{T.}~\bibnamefont{Vojta}},
  \bibinfo{journal}{Phys.Rev.Lett.} \textbf{\bibinfo{volume}{82}},
  \bibinfo{pages}{4707} (\bibinfo{year}{1999}).

\bibitem[{\citenamefont{Coleman et~al.}(2001)\citenamefont{Coleman, P\'epin,
  Si, and Ramazashvili}}]{col01a}
\bibinfo{author}{\bibfnamefont{P.}~\bibnamefont{Coleman}},
  \bibinfo{author}{\bibfnamefont{C.}~\bibnamefont{P\'epin}},
  \bibinfo{author}{\bibfnamefont{Q.}~\bibnamefont{Si}}, \bibnamefont{and}
  \bibinfo{author}{\bibfnamefont{R.}~\bibnamefont{Ramazashvili}},
  \bibinfo{journal}{J.Phys.:Condens.Mat.} \textbf{\bibinfo{volume}{13}},
  \bibinfo{pages}{R723} (\bibinfo{year}{2001}).

\bibitem[{\citenamefont{Si et~al.}(2001)\citenamefont{Si, Rabello, Ingersent,
  and Smith}}]{siq01a}
\bibinfo{author}{\bibfnamefont{Q.}~\bibnamefont{Si}},
  \bibinfo{author}{\bibfnamefont{S.}~\bibnamefont{Rabello}},
  \bibinfo{author}{\bibfnamefont{K.}~\bibnamefont{Ingersent}},
  \bibnamefont{and} \bibinfo{author}{\bibfnamefont{J.~L.} \bibnamefont{Smith}},
  \bibinfo{journal}{Nature} \textbf{\bibinfo{volume}{413}},
  \bibinfo{pages}{804} (\bibinfo{year}{2001}).

\bibitem[{\citenamefont{Hlubina and Rice}(1995)}]{hlu95a}
\bibinfo{author}{\bibfnamefont{R.}~\bibnamefont{Hlubina}} \bibnamefont{and}
  \bibinfo{author}{\bibfnamefont{T.~M.} \bibnamefont{Rice}},
  \bibinfo{journal}{Phys.Rev.B} \textbf{\bibinfo{volume}{51}},
  \bibinfo{pages}{9253} (\bibinfo{year}{1995}).

\bibitem[{\citenamefont{Rosch}(1999)}]{ros99a}
\bibinfo{author}{\bibfnamefont{A.}~\bibnamefont{Rosch}},
  \bibinfo{journal}{Phys.Rev.Lett.} \textbf{\bibinfo{volume}{82}},
  \bibinfo{pages}{4280} (\bibinfo{year}{1999}).

\bibitem[{\citenamefont{Rosch}(2000)}]{ros00a}
\bibinfo{author}{\bibfnamefont{A.}~\bibnamefont{Rosch}},
  \bibinfo{journal}{Phys.Rev.B} \textbf{\bibinfo{volume}{62}},
  \bibinfo{pages}{4945} (\bibinfo{year}{2000}).

\bibitem[{\citenamefont{Julian et~al.}(1998)\citenamefont{Julian, Carter,
  Grosche, Haselwimmer, Lister, Mathur, McMullan, Pfleiderer, Saxena, Walker
  et~al.}}]{jul98a}
\bibinfo{author}{\bibfnamefont{S.~R.} \bibnamefont{Julian}},
  \bibinfo{author}{\bibfnamefont{F.~V.} \bibnamefont{Carter}},
  \bibinfo{author}{\bibfnamefont{F.~M.} \bibnamefont{Grosche}},
  \bibinfo{author}{\bibfnamefont{R.~K.~W.} \bibnamefont{Haselwimmer}},
  \bibinfo{author}{\bibfnamefont{S.~J.} \bibnamefont{Lister}},
  \bibinfo{author}{\bibfnamefont{N.~D.} \bibnamefont{Mathur}},
  \bibinfo{author}{\bibfnamefont{G.~J.} \bibnamefont{McMullan}},
  \bibinfo{author}{\bibfnamefont{C.}~\bibnamefont{Pfleiderer}},
  \bibinfo{author}{\bibfnamefont{S.~S.} \bibnamefont{Saxena}},
  \bibinfo{author}{\bibfnamefont{I.~R.} \bibnamefont{Walker}},
  \bibnamefont{et~al.}, \bibinfo{journal}{J.Magn.Magn.Mat.}
  \textbf{\bibinfo{volume}{177-181}}, \bibinfo{pages}{265}
  (\bibinfo{year}{1998}).

\bibitem[{\citenamefont{Grosche et~al.}(2000)\citenamefont{Grosche, Steiner,
  Agarwal, Walker, Freye, Julian, and Lonzarich}}]{gro00a}
\bibinfo{author}{\bibfnamefont{F.~M.} \bibnamefont{Grosche}},
  \bibinfo{author}{\bibfnamefont{M.~J.} \bibnamefont{Steiner}},
  \bibinfo{author}{\bibfnamefont{P.}~\bibnamefont{Agarwal}},
  \bibinfo{author}{\bibfnamefont{I.~R.} \bibnamefont{Walker}},
  \bibinfo{author}{\bibfnamefont{D.~M.} \bibnamefont{Freye}},
  \bibinfo{author}{\bibfnamefont{S.~R.} \bibnamefont{Julian}},
  \bibnamefont{and} \bibinfo{author}{\bibfnamefont{G.~G.}
  \bibnamefont{Lonzarich}}, \bibinfo{journal}{Physica B}
  (\bibinfo{year}{2000}).

\bibitem[{\citenamefont{Bullett}(1982)}]{bul82a}
\bibinfo{author}{\bibfnamefont{D.~W.} \bibnamefont{Bullett}},
  \bibinfo{journal}{J.Phys.C} \textbf{\bibinfo{volume}{15}},
  \bibinfo{pages}{6163} (\bibinfo{year}{1982}).

\bibitem[{\citenamefont{Matsuura et~al.}(1998)\citenamefont{Matsuura, Watanabe,
  Kim, Doniach, Shen, Thio, and Bennett}}]{mat98b}
\bibinfo{author}{\bibfnamefont{A.~Y.} \bibnamefont{Matsuura}},
  \bibinfo{author}{\bibfnamefont{H.}~\bibnamefont{Watanabe}},
  \bibinfo{author}{\bibfnamefont{C.}~\bibnamefont{Kim}},
  \bibinfo{author}{\bibfnamefont{S.}~\bibnamefont{Doniach}},
  \bibinfo{author}{\bibfnamefont{Z.-X.} \bibnamefont{Shen}},
  \bibinfo{author}{\bibfnamefont{T.}~\bibnamefont{Thio}}, \bibnamefont{and}
  \bibinfo{author}{\bibfnamefont{J.~W.} \bibnamefont{Bennett}},
  \bibinfo{journal}{Phys.Rev.B} \textbf{\bibinfo{volume}{58}},
  \bibinfo{pages}{3690} (\bibinfo{year}{1998}).

\bibitem[{\citenamefont{Wilson and Pitt}(1971)}]{wil71a}
\bibinfo{author}{\bibfnamefont{J.~A.} \bibnamefont{Wilson}} \bibnamefont{and}
  \bibinfo{author}{\bibfnamefont{G.~D.} \bibnamefont{Pitt}},
  \bibinfo{journal}{Philosoph.Mag.} \textbf{\bibinfo{volume}{23}},
  \bibinfo{pages}{1297} (\bibinfo{year}{1971}).

\bibitem[{\citenamefont{Sekine et~al.}(1997)\citenamefont{Sekine, Takahashi,
  Mori, Matsumoto, and Kosaka}}]{sek97a}
\bibinfo{author}{\bibfnamefont{Y.}~\bibnamefont{Sekine}},
  \bibinfo{author}{\bibfnamefont{H.}~\bibnamefont{Takahashi}},
  \bibinfo{author}{\bibfnamefont{N.}~\bibnamefont{Mori}},
  \bibinfo{author}{\bibfnamefont{T.}~\bibnamefont{Matsumoto}},
  \bibnamefont{and} \bibinfo{author}{\bibfnamefont{T.}~\bibnamefont{Kosaka}},
  \bibinfo{journal}{Physica B} \textbf{\bibinfo{volume}{237-238}},
  \bibinfo{pages}{148} (\bibinfo{year}{1997}).

\bibitem[{\citenamefont{Wilson}(1972)}]{wil72a}
\bibinfo{author}{\bibfnamefont{J.~A.} \bibnamefont{Wilson}},
  \bibinfo{journal}{Adv. in Phys.} \textbf{\bibinfo{volume}{21}},
  \bibinfo{pages}{143} (\bibinfo{year}{1972}).

\bibitem[{\citenamefont{Wilson}(1997)}]{wil85a}
\bibinfo{author}{\bibfnamefont{J.~A.} \bibnamefont{Wilson}}, in
  \emph{\bibinfo{booktitle}{The Metallic and Non-Metallic States of Matter}},
  edited by \bibinfo{editor}{\bibfnamefont{P.~P.} \bibnamefont{Edwards}}
  \bibnamefont{and} \bibinfo{editor}{\bibfnamefont{C.~N.~R.} \bibnamefont{Rao}}
  (\bibinfo{publisher}{Taylor and Francis}, \bibinfo{address}{London},
  \bibinfo{year}{1997}), chap.~\bibinfo{chapter}{9}, pp.
  \bibinfo{pages}{215--260}.

\bibitem[{\citenamefont{Mori and Takahashi}(1983)}]{mor83a}
\bibinfo{author}{\bibfnamefont{N.}~\bibnamefont{Mori}} \bibnamefont{and}
  \bibinfo{author}{\bibfnamefont{H.}~\bibnamefont{Takahashi}},
  \bibinfo{journal}{J.Magn.Magn.Mat} \textbf{\bibinfo{volume}{31-34}},
  \bibinfo{pages}{335} (\bibinfo{year}{1983}).

\bibitem[{\citenamefont{Husmann et~al.}(2000)\citenamefont{Husmann, Brooke,
  Rosenbaum, Yao, and Honig}}]{hus00a}
\bibinfo{author}{\bibfnamefont{A.}~\bibnamefont{Husmann}},
  \bibinfo{author}{\bibfnamefont{J.}~\bibnamefont{Brooke}},
  \bibinfo{author}{\bibfnamefont{T.~F.} \bibnamefont{Rosenbaum}},
  \bibinfo{author}{\bibfnamefont{X.}~\bibnamefont{Yao}}, \bibnamefont{and}
  \bibinfo{author}{\bibfnamefont{J.~M.} \bibnamefont{Honig}},
  \bibinfo{journal}{Phys.Rev.Lett.} \textbf{\bibinfo{volume}{84}},
  \bibinfo{pages}{2465} (\bibinfo{year}{2000}).

\bibitem[{\citenamefont{Wittig}(1966)}]{wit66a}
\bibinfo{author}{\bibfnamefont{J.}~\bibnamefont{Wittig}},
  \bibinfo{journal}{Z.Phys.} \textbf{\bibinfo{volume}{195}},
  \bibinfo{pages}{215} (\bibinfo{year}{1966}).

\bibitem[{\citenamefont{Eichler and Wittig}(1968)}]{eic68a}
\bibinfo{author}{\bibfnamefont{A.}~\bibnamefont{Eichler}} \bibnamefont{and}
  \bibinfo{author}{\bibfnamefont{J.}~\bibnamefont{Wittig}},
  \bibinfo{journal}{Z.Angew.Phys.} \textbf{\bibinfo{volume}{25}},
  \bibinfo{pages}{319} (\bibinfo{year}{1968}).

\bibitem[{\citenamefont{Yao et~al.}(1996)\citenamefont{Yao, Honig, Hogan,
  Kannewurf, and Spalek}}]{yao96a}
\bibinfo{author}{\bibfnamefont{X.}~\bibnamefont{Yao}},
  \bibinfo{author}{\bibfnamefont{J.~M.} \bibnamefont{Honig}},
  \bibinfo{author}{\bibfnamefont{T.}~\bibnamefont{Hogan}},
  \bibinfo{author}{\bibfnamefont{C.}~\bibnamefont{Kannewurf}},
  \bibnamefont{and} \bibinfo{author}{\bibfnamefont{J.}~\bibnamefont{Spalek}},
  \bibinfo{journal}{Phys.Rev.B} \textbf{\bibinfo{volume}{54}},
  \bibinfo{pages}{17469} (\bibinfo{year}{1996}).

\bibitem[{\citenamefont{Sudo and Miyadai}(1985)}]{sud85a}
\bibinfo{author}{\bibfnamefont{S.}~\bibnamefont{Sudo}} \bibnamefont{and}
  \bibinfo{author}{\bibfnamefont{T.}~\bibnamefont{Miyadai}},
  \bibinfo{journal}{J.Phys.Soc.Jpn.} \textbf{\bibinfo{volume}{54}},
  \bibinfo{pages}{3934} (\bibinfo{year}{1985}).

\bibitem[{\citenamefont{Sudo}(1992)}]{sud92a}
\bibinfo{author}{\bibfnamefont{S.}~\bibnamefont{Sudo}},
  \bibinfo{journal}{J.Magn.Magn.Mat.} \textbf{\bibinfo{volume}{114}},
  \bibinfo{pages}{57} (\bibinfo{year}{1992}).

\bibitem[{\citenamefont{Hauser et~al.}(1995)\citenamefont{Hauser, Indinger,
  Bauer, and Gratz}}]{hau95a}
\bibinfo{author}{\bibfnamefont{R.}~\bibnamefont{Hauser}},
  \bibinfo{author}{\bibfnamefont{A.}~\bibnamefont{Indinger}},
  \bibinfo{author}{\bibfnamefont{E.}~\bibnamefont{Bauer}}, \bibnamefont{and}
  \bibinfo{author}{\bibfnamefont{E.}~\bibnamefont{Gratz}},
  \bibinfo{journal}{J.Magn.Magn.Mat.} \textbf{\bibinfo{volume}{140-144}},
  \bibinfo{pages}{799} (\bibinfo{year}{1995}).

\bibitem[{\citenamefont{Pfleiderer et~al.}(1997)\citenamefont{Pfleiderer,
  McMullan, Julian, and Lonzarich}}]{pfl97a}
\bibinfo{author}{\bibfnamefont{C.}~\bibnamefont{Pfleiderer}},
  \bibinfo{author}{\bibfnamefont{G.~J.} \bibnamefont{McMullan}},
  \bibinfo{author}{\bibfnamefont{S.~R.} \bibnamefont{Julian}},
  \bibnamefont{and} \bibinfo{author}{\bibfnamefont{G.~G.}
  \bibnamefont{Lonzarich}}, \bibinfo{journal}{Phys.Rev.B}
  \textbf{\bibinfo{volume}{55}}, \bibinfo{pages}{8330} (\bibinfo{year}{1997}).

\bibitem[{\citenamefont{Thessieu et~al.}(1997)\citenamefont{Thessieu,
  Pfleiderer, Stepanov, and Flouquet}}]{the97a}
\bibinfo{author}{\bibfnamefont{C.}~\bibnamefont{Thessieu}},
  \bibinfo{author}{\bibfnamefont{C.}~\bibnamefont{Pfleiderer}},
  \bibinfo{author}{\bibfnamefont{A.~N.} \bibnamefont{Stepanov}},
  \bibnamefont{and} \bibinfo{author}{\bibfnamefont{J.}~\bibnamefont{Flouquet}},
  \bibinfo{journal}{J.Phys.Cond.Mat.} \textbf{\bibinfo{volume}{9}},
  \bibinfo{pages}{6677} (\bibinfo{year}{1997}).

\bibitem[{nik()}]{nik06a}
\bibinfo{note}{The order of magnitude, $T_{sf}$, was estimated from the value
  of $A$ in the SCR model, $k_F$ from the size of the Brillouin zone and
  electron filling of the bands and $l$ from $\rho_0$.}

\end{thebibliography}

\end{document}